\documentclass{article}
\usepackage[utf8]{inputenc}
\usepackage{authblk}
\usepackage{setspace}
\usepackage[margin=1.25in]{geometry}
\usepackage{graphicx}
\graphicspath{ {./figures/} }
\usepackage{subcaption}
\usepackage{amsmath}
\usepackage{lineno}
\usepackage{siunitx}

\usepackage[style=nejm, 
citestyle=numeric-comp,
sorting=none]{biblatex}
\title{AttoSHINE: Generation of continuous-wave terawatt-scale attosecond X-ray pulses at SHINE}

\author[1,2*]{Bingyang Yan}
\author[1,2,5*]{Chenzhi Xu}
\author[3]{Si Chen}
\author[3]{Duan Gu}
\author[4]{Ye Chen}
\author[4$\dag$]{Jiawei Yan}
\author[3$\ddagger$]{Haixiao Deng}

\affil[1]{Shanghai Institute of Applied Physics, Chinese Academy of Sciences, Shanghai 201800, China.}
\affil[2]{University of Chinese Academy of Sciences, Beijing 100049, China.}
\affil[3]{Shanghai Advanced Research Institute, Chinese Academy of Sciences, Shanghai 201210, China.}
\affil[4]{Deutsches Elektronen-Synchrotron DESY, 22603 Hamburg, Germany}
\affil[5]{European XFEL, Schenefeld 22869, Germany.}

\affil[*]{These authors have contributed equally to this work.}
\affil[$\dag$]{Address correspondence to: jiawei.yan@desy.de}
\affil[$\ddagger$]{Address correspondence to: denghx@sari.ac.cn}

\date{}

\begin{document}

\maketitle

\begin{abstract}
Attosecond X-ray pulses are a critical tool for tracking ultrafast electron dynamics in condensed matter, molecular systems, and strongly correlated materials. Recent breakthroughs have pushed X-ray free electron lasers (XFELs) into the attosecond domain, significantly surpassing their previous femtosecond capabilities. Building on these advancements, this work investigates the potential of the Shanghai HIgh repetitioN rate XFEL and Extreme light facility (SHINE), China's first continuous-wave (CW) XFEL, to generate intense attosecond X-ray pulses, thereby offering transformative capabilities for X-ray science. Through comprehensive start-to-end simulations, we show that SHINE is capable of producing terawatt-scale pulses at both soft and hard X-ray ranges. This is achieved using a self-chirping scheme within the existing machine configuration, requiring no additional hardware. We further find that superradiant behavior in the postsaturation regime plays a central role in attosecond XFEL, enabling higher peak power and shorter attosecond pulses. Our results illustrate the wide applicability of the self-chirping scheme and demonstrate that CW XFELs can generate intense attosecond X-ray pulses at megahertz repetition rates, opening new opportunities for real-time studies of electronic dynamics in complex systems.

\end{abstract}


\section{Introduction}

Attosecond--duration, \AA ngstrom--wavelength X-ray pulses match the intrinsic temporal and spatial scales of electron motion in solids and molecules, making them indispensable for resolving the fundamental processes that govern matter~\cite{aaa2}. In the past two decades, high-harmonic generation (HHG) \cite{hhg1, hhg2, hhg3,hhg4,hhg5} has been the dominant technique for producing attosecond pulses in the extreme ultraviolet and soft X-ray regimes \cite{hhgsoft1, hhgsoft2,hhgsoft3}. However, extending HHG to the \AA ngstrom--wavelength remains highly challenging due to the rapidly decreasing conversion efficiency at higher photon energies.

In parallel, X-ray free-electron lasers (XFELs) \cite{fel1,fel2,fel3,fel4,fel5,fel6} have opened unprecedented opportunities in X-ray science due to their capabilities of generating high-brightness X-ray pulses covering a wide range of photon energies. These facilities predominantly operate based on the self-amplified spontaneous emission (SASE) process \cite{sase1,sase2}. In this mechanism, a relativistic electron beam traversing a periodic magnetic structure (undulator) undergoes a strong interaction with its own spontaneous emission electromagnetic field, leading to the exponential amplification of the initial radiation starting from shot noise within the electron beam. This enables modern XFEL facilities to deliver X-ray pulses with peak brightness orders of magnitude beyond those of tabletop sources. 

XFEL facilities worldwide have routinely delivered intense femtosecond pulses. More recently, significant progress has been made toward the generation of attosecond XFEL pulses. In the soft X-ray regime, a variety of techniques have been explored and experimentally demonstrated, including self-modulation in long-period wigglers \cite{soft1}, cathode shaping \cite{soft2,soft5}, chirp–dispersion scheme~\cite{serkez:fel2022-tuai2,funke2024capturing}, laser heater shaping \cite{soft4}, and low-charge operation \cite{soft3}. In the hard X-ray regime, attosecond pulse generation has been realized using low-charge operation at LCLS \cite{PhysRevSTAB.17.120703,hard2,hard3} and SwissFEL \cite{hard4}, achieving pulse energies in the range of a few to tens of microjoules. More recently, a self-chirping scheme was proposed and demonstrated at the European XFEL to generate terawatt-scale hard X-ray attosecond pulses \cite{hard}, significantly advancing their experimental usability \cite{zhu2024attosecond}. Unlike previous approaches, the self-chirping scheme requires no additional hardware (such as the wiggler used for self-modulation or cathode shaping schemes \cite{soft1,soft2}) and does not rely on ultra-low bunch charge, enabling straightforward adoption at existing XFELs that include a dogleg or arc section and thereby greatly improving the availability of attosecond XFEL pulses. Notably, XFELs remain the only sources currently capable of producing intense hard X-ray attosecond pulses.

The advent of continuous-wave (CW) XFELs \cite{cwlcls2015,cwlcls2021,cwshine2024,cwshine2023,cws3fel, zhu2022inhibition}, enabled by superconducting linear accelerators operating at megahertz (MHz) repetition rates, marks a transformative development in the field. These next-generation light sources significantly enhance the average spectral brightness and unlock new experimental frontiers in areas such as X-ray crystallography, single-particle imaging, and resonant inelastic X-ray scattering. The realization of attosecond pulses in a CW XFEL could enable new capabilities, including electronic damage-free structural measurements \cite{chapman2025convergent, dam1, dam2} and access to nonlinear X-ray phenomena at unprecedented repetition rates. However, the feasibility of implementing advanced attosecond-generation techniques, particularly the self-chirping scheme, within CW XFELs driven by normal conducting very-high-frequency (VHF) photoinjectors remains an open and critical question. A comprehensive understanding of both the practical implementation and the fundamental performance limits of CW attosecond XFELs is therefore essential.

This paper explains the universality of the self-chirping scheme through presenting the first design study of a CW attosecond XFEL based on the parameters of the Shanghai HIgh repetitioN rate XFEL and Extreme light facility (SHINE) \cite{shine1,cwshine2024,yan2019multi}. Section II introduces the physical principles underlying the self-chirping scheme and outlines the required modifications to the SHINE facility. Section III presents beam dynamics simulations and the stability analysis under the CW mode, and introduces a new case for the tail current spike. Section IV provides FEL simulation results demonstrating the generation of sub-femtosecond X-ray pulses. Section V concludes with a summary and outlook.

\section{Self-chirping based AttoSHINE proposal}

In typical XFEL operation, the desired high peak current profile is achieved by adjusting the phases of the radio-frequency (rf) structures before the magnetic chicane. For a uniform final current profile, the energy chirp before compression should be linear. This is commonly accomplished using a harmonic rf structure, often referred to as a longitudinal phase space linearizer, which cancels the nonlinear rf curvature by decelerating the beam appropriately. Moreover, two-frequency rf systems in conjunction with bunch compressors can also be used to manipulate the nonlinear correlations and generate ramped current profiles. Previous studies have shown that adjusting the linac phase and then compressing the beam through a chicane produces a banana-shaped bunch with a high peak current at head \cite{ackermann2007operation,hard2,beamshape1}.

\begin{figure}
\centering
\includegraphics[width=0.95\linewidth]{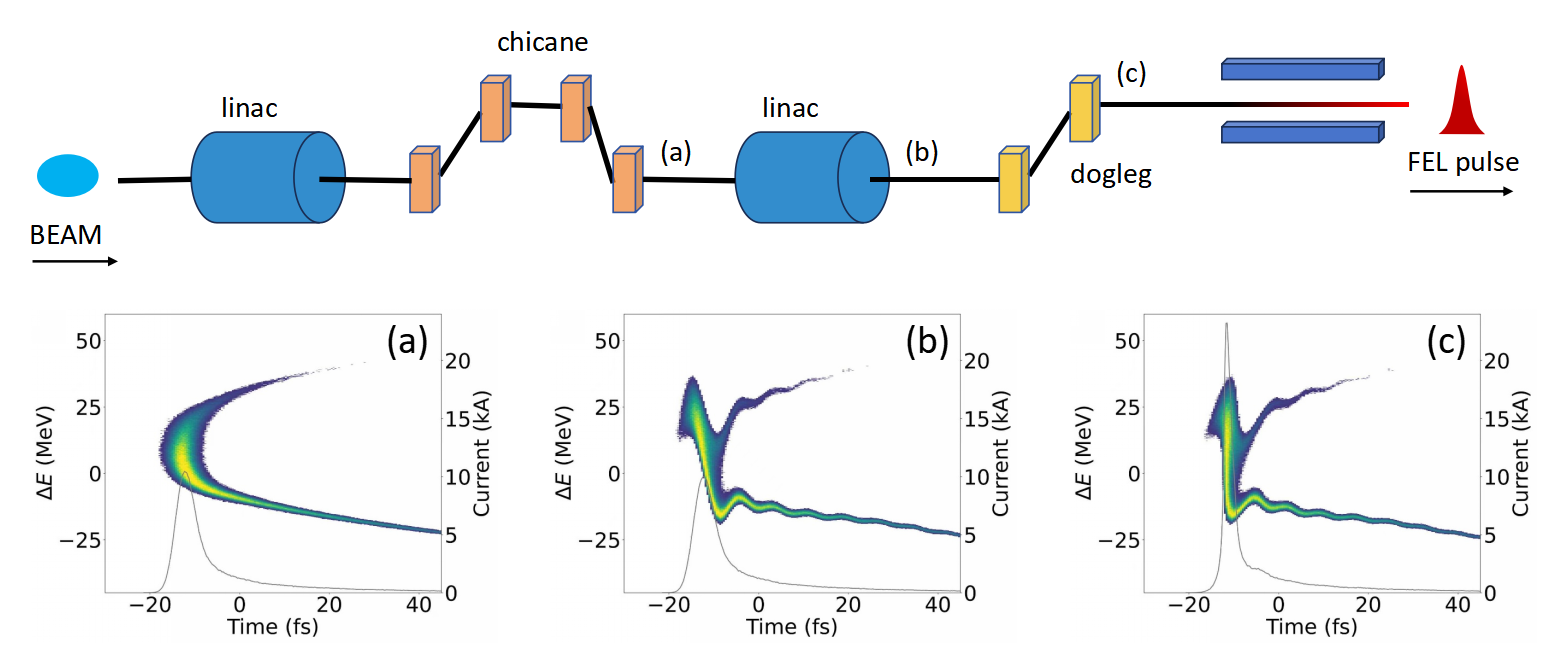}
\caption{ Schematic illustration of the self-chirping scheme, which includes two linac sections, a chicane, a dogleg section, and an undulator. (a)-(c) show the evolution of the longitudinal phase space and corresponding current profiles (gray lines). Bunch head is to the left.}
\label{fig:01}
\end{figure}

The key physical principle underlying the self-chirping scheme is to shape the electron bunch into a sharp current profile within the linac and exploit collective effects to induce an energy chirp. This chirp is then utilized to achieve strong compression of the electron beam immediately before it enters the undulator. Figure \ref{fig:01} illustrates a typical self-chirping scheme. A high-order curvature is introduced into the beam during acceleration in the rf structure. After passing through a bunch compression section, a high-current region develops at the head of the beam. As the beam propagates downstream, the longitudinal space charge (LSC) effect \cite{LSC}, a collective effect arising from the charge distribution, induces a self-generated energy chirp in this high-current region. The self-chirped beam is then transported through a dogleg or arc composed of double-bend achromats (DBAs). In conventional beamlines, such dogleg or arc sections are typically designed to be isochronous with $R_{56}=0$ \cite{SCLASW, shineswitchyard}, or to have only a small residual $R_{56}$ \cite{wang2011optimization,hard2}, resulting in only weak longitudinal phase space rotation of a chirped bunch. In contrast, the self-chirping scheme intentionally employs a dogleg or arc with a large positive $R_{56}$ to enable strong compression to the $\sim$1~fs scale. The degree of bunch compression is determined by the combined effect of the beam energy chirp and the transport matrix element $R_{56}$, which describes the longitudinal displacement induced by a relative energy deviation \cite{BC}. Unlike low-charge nonlinear-compression modes \cite{PhysRevSTAB.17.120703, hard2, hard4, soft3}, the self-chirping approach does not sacrifice bunch charge. Rather than using nonlinear compression to directly generate a short current spike, it serves only to imprint a large energy chirp, which is then converted into an ultra-high peak current spike by the tailored $R_{56}$ of the dogleg or arc. As a result, the self-chirping scheme emphasizes efficient bunch compression immediately before the undulator.

\begin{figure}
\centering 
\includegraphics[width=0.95\linewidth]{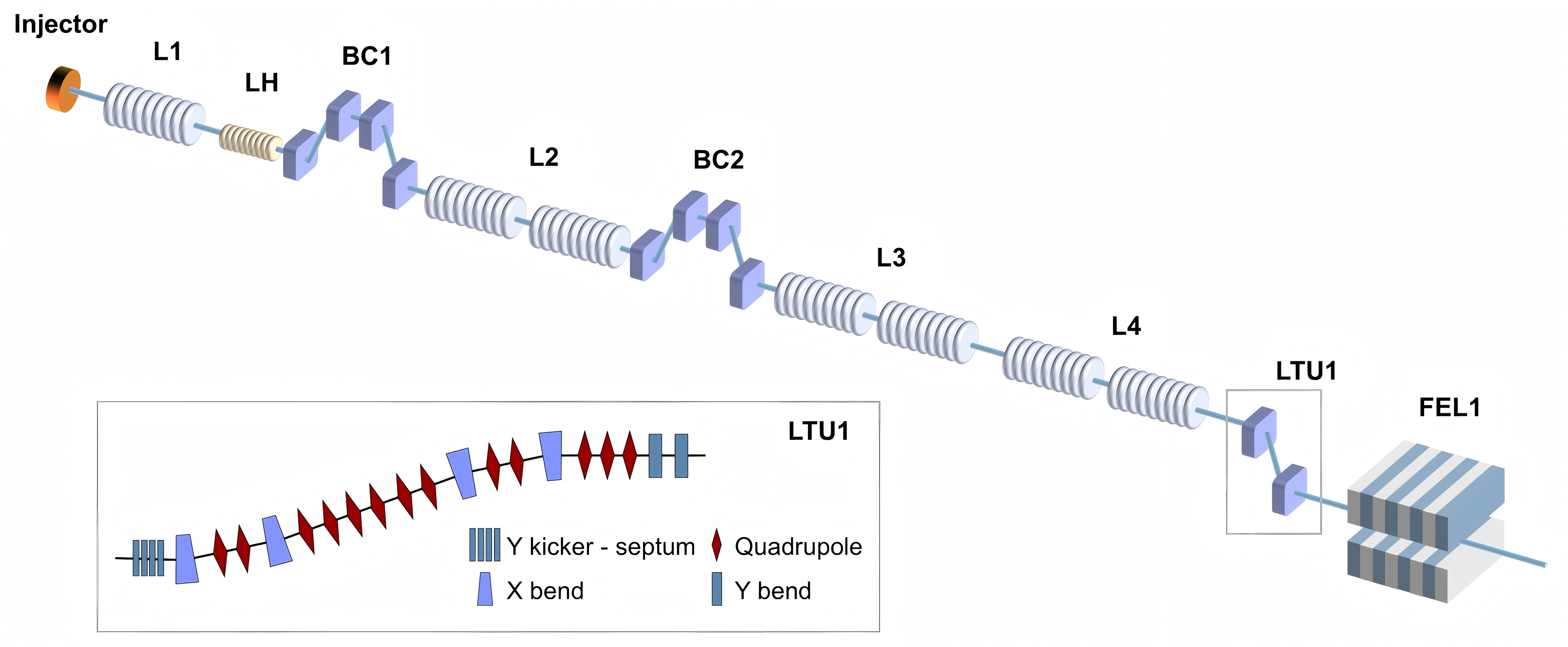}
\caption{ Layout of AttoSHINE with LTU1 beamline shown as a representative beamline section. Here, L1, LH, L2, and L3 denote linac sections, BC1 and BC2 are bunch compression sections, and FEL1 refers to the undulator section. The grey box highlights the layout of the deflection section of the LTU1 beamline.}
\label{fig:2}
\end{figure}

In the self-chirping scheme, the LSC interaction is one of the main sources of the imposed energy chirp. In an ideal conducting cylindrical pipe with a beam of uniform circular cross section, the longitudinal field on axis produced by the LSC effect is \cite{LSC}
\[
E_z(z)\;=\;
-\,\frac{1}{2\pi\,\varepsilon_{0}\,\gamma^{2}}
\left(\ln\frac{b}{a}+\frac{1}{2}\right)
\;\frac{d\lambda(z)}{dz},
\]
where $\varepsilon_{0}$ is the vacuum permittivity, $\gamma$ is the electron energy in units of its rest mass $mc^{2}$, $a$ is the radius of the transverse cross section for a uniform distribution and can be approximately taken as the sum of rms transverse beam sizes for a Gaussian or a parabolic distribution, $b$ is the radius of the round pipe, and $\lambda(z)$ is the line charge density of the beam in the laboratory frame. Because the LSC field scales with the axial derivative of the line current, electrons located ahead of the current peak gain energy, whereas those behind it lose energy, producing the desired energy modulation. Since the LSC impedance falls off as $1/\gamma^{2}$, this current profile shaping should be performed at low beam energy. By appropriately controlling the peak current, the amplitude of this chirp can be tuned. In addition to LSC, coherent synchrotron radiation (CSR) \cite{csr} inevitably arises as the electron beam passes through magnetic chicanes and dogleg sections. The CSR field, emitted during bending, can further enhance the energy chirp, particularly in high-current regions. Moreover, the CSR-induced energy modulation couples with dispersion in the bending sections, leading to off-axis displacement of the beam \cite{ob,dt}. Therefore, CSR should be carefully accounted for in the simulation, and it is essential to optimize the electron beam orbit in the undulator to ensure that the high-current portion remains aligned on-axis.

At the European XFEL \cite{hard}, the self-chirping concept is implemented with an arc section that compresses the self-chirped beam. However, increasing the $R_{56}$ through quadrupole optimization in the arc also introduces transverse dispersion. While the transverse dispersion can help suppress lasing from the low-current portion of the bunch and shorten the output pulse duration, it also poses additional challenges for orbit control of the electron beam in the undulator. To this end, we aim to maximize the $R_{56}$ for final bunch compression while simultaneously minimizing transverse dispersion leakage downstream of the arc.

Building on this approach, we propose to generate attosecond XFEL pulses at SHINE based on the self-chirping concept, hereafter referred to as AttoSHINE. The injector can generate 1 MHz CW bunches and accelerate them to 100 MeV. These beams are then typically accelerated to 8 GeV through four linear accelerator sections and compressed to 1500 A by a selectable two-stage or three-stage compression scheme at SHINE. In the beam switchyard section \cite{shineswitchyard}, the bunches are kicked by fast kickers into two beam transport lines (LTU1, LTU2), which deliver them to two downstream undulator sections (FEL1, FEL2), respectively. The layout of the AttoSHINE and the deflection section of the LTU1 beamline is shown in figure \ref{fig:2}. 

\begin{figure}
\centering
\includegraphics[width=0.95\linewidth]{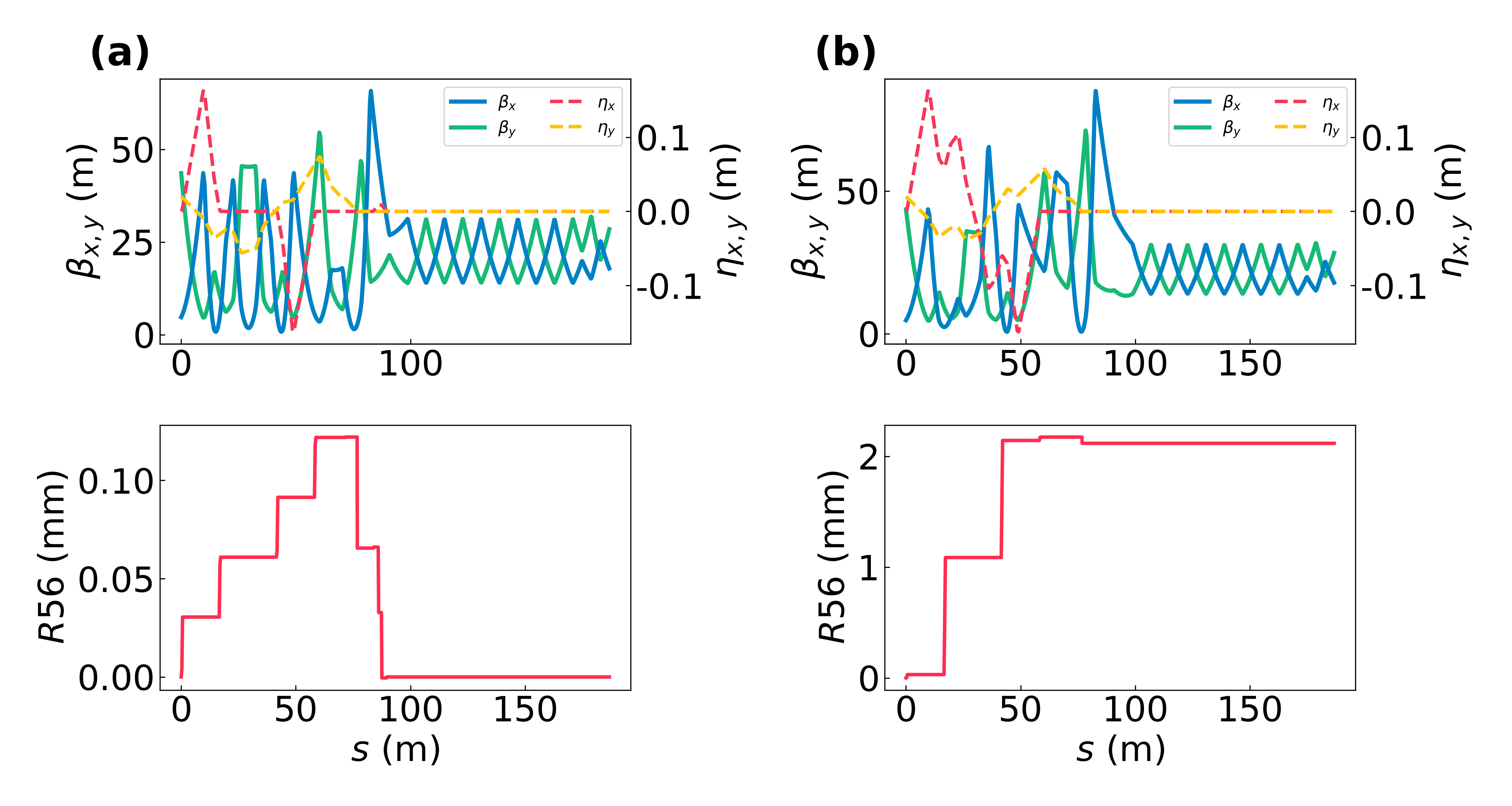}
\caption{ Optical functions and $R_{56}$ along the LTU1 beamline. (a) the original design; (b) an optimized lattice with a $R_{56}$ of 2.1 mm while remaining achromatic.}
\label{fig:lattice}
\end{figure}

The self-chirping process is initiated by appropriately tuning the accelerating phases of linac sections L1, L2 and the third-harmonic rf section (LH). And then, the compression process is based on the LTU1 beamline. A kicker–septum combination first extracts the beam from the straight focusing-defocusing (FODO) lattice. The beam is then deflected by a horizontal dogleg consisting of two DBAs, which introduces the required horizontal offset while keeping the horizontal dispersion zero downstream of the second DBA. A subsequent vertical dogleg returns the beam to its original vertical plane and simultaneously cancels the vertical dispersion. In nominal operation, isochronicity is ensured by a downstream chicane, which is deactivated in the self‑chirping operation. Instead, the horizontal dogleg section provides a positive $R_{56}$ through adjustments of the quadrupole magnets. By relaxing the phase-advance requirements for optics balance, $R_{56}$ can be tuned flexibly while maintaining zero transverse dispersion in the downstream beamline. As shown in figure \ref{fig:lattice}, a $R_{56}$ of 2.1 mm is achievable while maintaining a dispersion-free condition in both transverse planes. This flexibility allows for the implementation of different transport schemes, thereby enabling different compression strategies of self-chirped bunches with varying characteristics. 

\section{Electron beam simulation results}

The SHINE design parameters were optimized to investigate the self-chirping scheme for generating attosecond X-ray FEL pulses. In the injector section \cite{chen22,zhu2022inhibition}, beam dynamics were simulated using ASTRA \cite{astra} with one million macroparticles, while downstream linac and transport line were simulated using ELEGANT \cite{elegant}. Collective effects including the LSC, CSR, and wakefields were included in the simulation. Here, the two-stage compression scheme is used for the self-chirping scheme. The beam passes through five linac sections (L1, LH, L2, L3 and L4) and two bunch compression sections (BC1 and BC2). The LH section includes a 3.9 GHz third harmonic rf structure while the other accelerating sections are composed of 1.3 GHz cryomodules. After BC2, linac sections L3 and L4 provide further acceleration and are operated at zero phase in order to avoid perturbations to the longitudinal phase space, raising the beam energy to around 8 GeV. The phases of the L1, LH, and L2 cavities were tuned to optimize the longitudinal phase space and to realize the beam‑shaping after the BC2.

\begin{figure}
\centering
\includegraphics[width=1\linewidth]{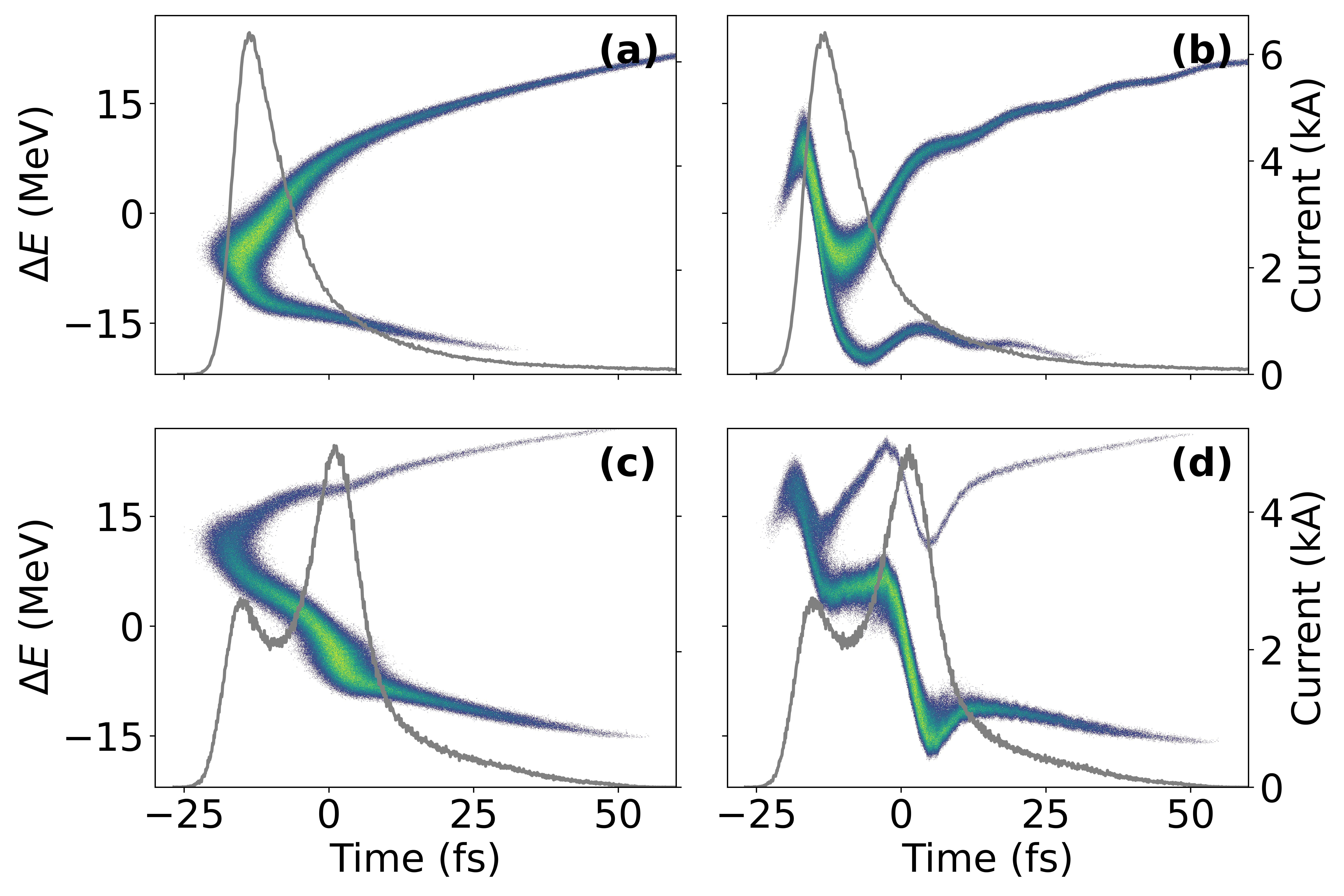}
\caption{Longitudinal phase spaces of the head-current spike case (a, b) and tail-current spike case (c, d). The left corresponds to the end of BC2; the right corresponds to the end of L4. Grey lines represent the current profile. Bunch head is to the left.}
\label{fig:left}
\end{figure}

A ramped, high peak current profile is favorable for the self-chirping scheme, and is typically located at the head of the bunch, as shown in Ref.~\cite{hard}. In this study, we found that it is possible to tune the location of the current spike along the electron bunch, either near the head or closer to the tail. Figure \ref{fig:left} shows two representative cases obtained in our simulations, referred to as the head-current spike and tail-current spike cases, respectively. Figure \ref{fig:left}(a) and (b) show the longitudinal phase space distributions at the end of BC2 and L4 for the head-current spike case. In this working point, the L1 phase is set to $-4.6^\circ$, the LH phase to $-142.2^\circ$, and the L2 phase to $-33.4^\circ$. To optimize the slice energy spread at the entrance of the undulator, the peak current of the head spike after the BC2 was intentionally reduced to control the self-chirping process. After the BC2, the bunch exhibits a full width at half maximum (FWHM) length of approximately 10 fs and a peak current of 6 kA. 

Figure \ref{fig:left}(c) and (d) present the longitudinal phase space distribution after BC2 and L4 for the tail-current spike case. In this scenario, the L1 phase is adjusted to $-12.1^\circ$, the LH phase is set to $-152.0^\circ$, and the L2 phase is tuned to $-28.9^\circ$. The primary distinction of the tail spike case is that the current spike is located at the trailing edge of the electron bunch. After BC2, the longitudinal phase space distribution exhibits a small protrusion beneath the characteristic ``C shaped'' profile, where electrons accumulate and form a current spike. During subsequent transport through sections L3 and L4, this current spike region develops a sufficiently strong chirp due to the LSC effect. Additionally, a high-current region formed by nonlinear compression is also present at the bunch head, similarly contributing a minor chirp. Compared to the head-current spike case, the peak current in the tail-current spike case is lower, approximately 5 kA, resulting in a correspondingly smaller chirp. Moreover, between the two high-current regions lies a plateau region with a current of approximately 2 kA. 

\begin{figure}
\centering
\includegraphics[width=1\linewidth]{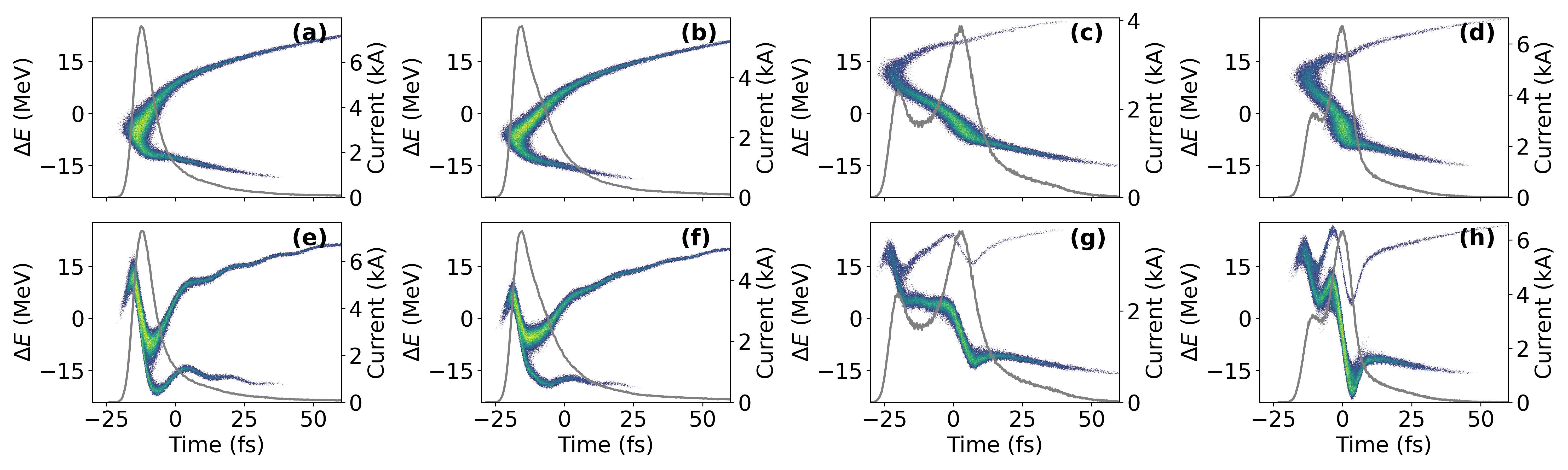}
\caption{Longitudinal phase space and current profile of different settings. (a)-(d) refer to the end of BC2: (a) head-current spike case with LH + $0.1^\circ$, (b) head-current spike case with LH - $0.1^\circ$, (c) tail-current spike case with LH + $0.1^\circ$, and (d) tail-current spike case with LH - $0.1^\circ$. (e)-(h) refer to the end of L4: (e) head-current spike case with LH + $0.1^\circ$, (f) head-current spike case with LH - $0.1^\circ$, (g) tail-current spike case with LH + $0.1^\circ$, and (h) tail-current spike case with LH - $0.1^\circ$. Bunch head is to the left.}
\label{fig:24}
\end{figure}

By adjusting the phase of individual linac sections, the longitudinal phase space distribution of the electron bunch can be precisely manipulated, allowing for fine control of the peak current. Figure \ref{fig:24} illustrates how variations in the LH rf phase affect the longitudinal phase space distribution in both the head- and tail-current spike cases. During these variations, the bunch compression factor is held constant, while the downstream magnet strengths are correspondingly adjusted to maintain the electron bunch energy at the central reference energy of the magnets. In the head-current spike case, increasing (or decreasing) the LH phase by $0.1^\circ$ results in a corresponding increase (or decrease) in the peak current by approximately 1 kA. In contrast, the tail-current spike case exhibits the opposite behavior, where a $0.1^\circ$ increase or decrease in the LH phase results in a decrease or increase of the peak current by about $1\,\mathrm{kA}$. These trends are primarily driven by changes in the first-order energy chirp induced by the linac rf phase. This chirp not only governs the final peak current and bunch length after compression, but in the tail-current spike case, it also significantly reshapes the overall current profile.

\begin{figure}
\includegraphics[width=1\linewidth]{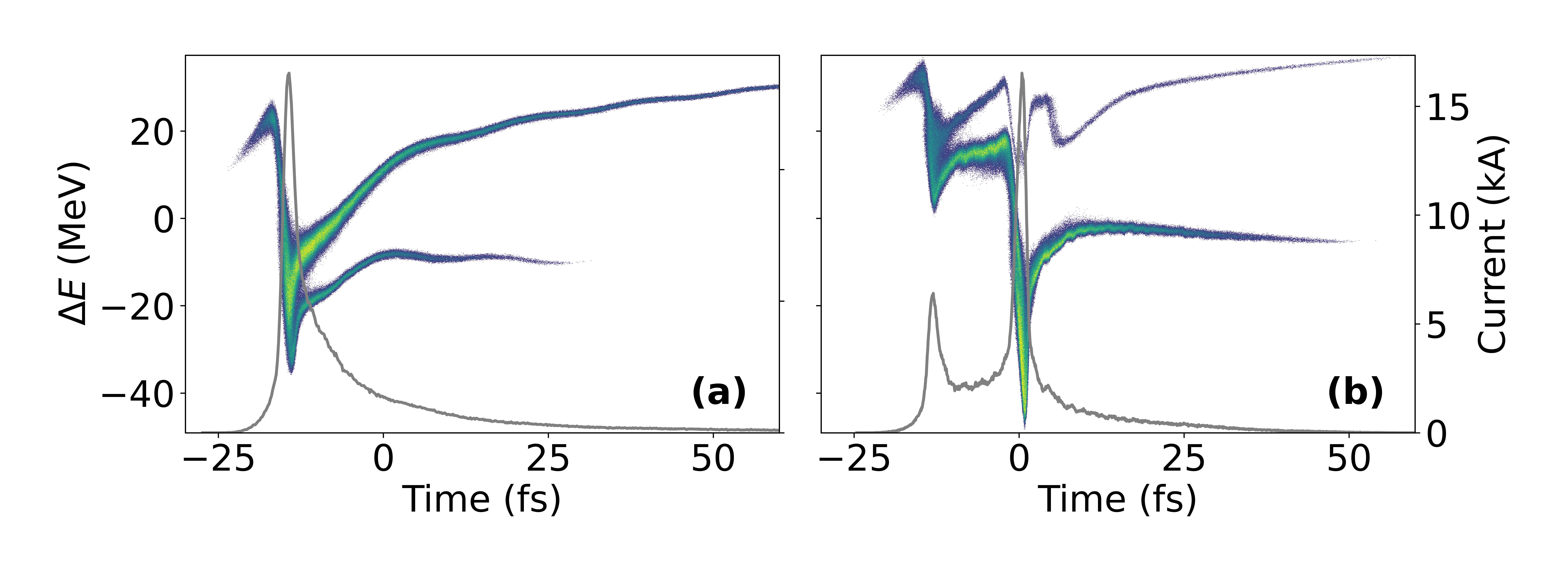}
\caption{Longitudinal phase space at the end of LTU1 beamline of the head-current spike case (a) and tail-current spike case (b). Bunch head is to the left.}
\label{fig:ltu1}
\end{figure}

\begin{figure}
\includegraphics[width=1\linewidth]{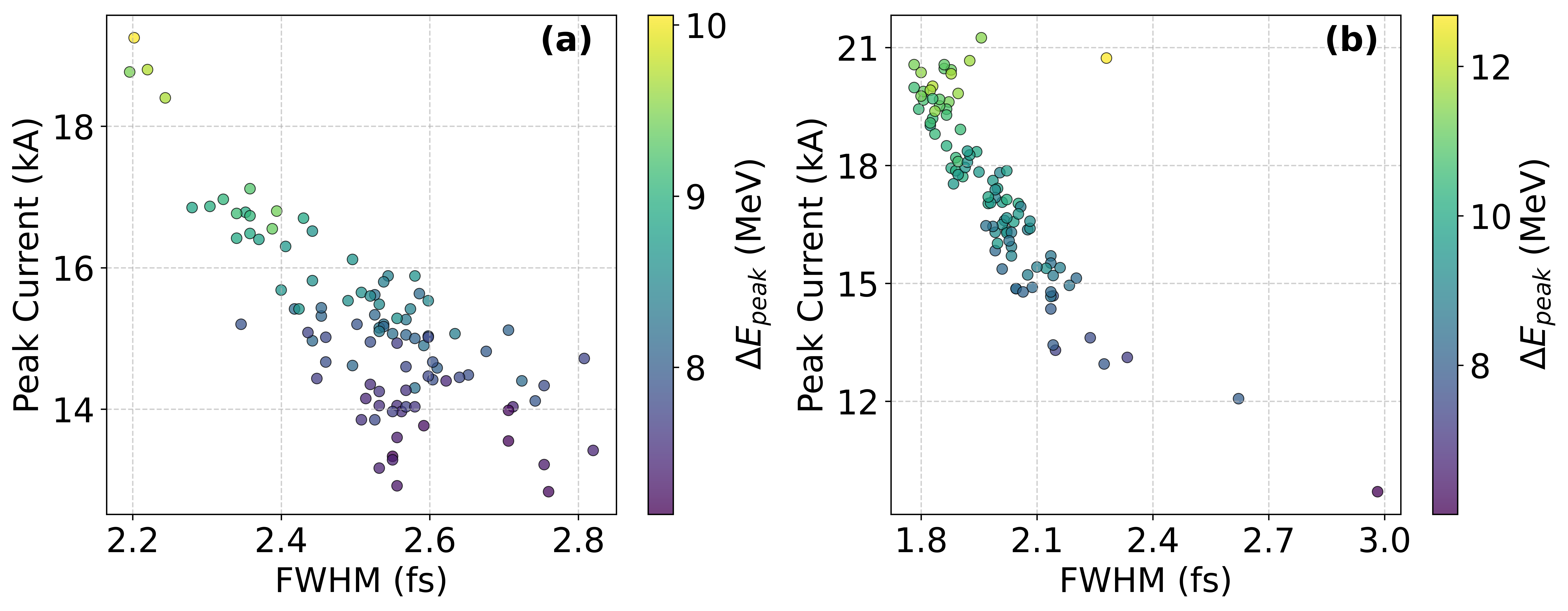}
\caption{Scatterplots of beam parameters from 100 simulations under specified rf phase and amplitude jitters with the head-current spike case (a) and the tail-current spike case (b).}
\label{fig:rms}
\end{figure}

Figure~\ref{fig:ltu1} shows the longitudinal phase space distributions of the electron bunch after the LTU1 beamline. To achieve compression, the \( R_{56} \) of LTU1 is adjusted according to the energy chirp of the bunch. For the head-current spike case, \( R_{56} = 0.38\,\mathrm{mm} \) yields a peak current of approximately 14~kA, with a slice energy spread of 8~MeV and a FWHM duration of 2.57~fs. At peak current, the horizontal slice emittance is 0.30 mm·mrad and the vertical slice emittance is 0.35 mm·mrad. For the tail-current spike case, a larger \( R_{56} = 0.60\,\mathrm{mm} \) is required to achieve the same level of compression due to the smaller energy chirp. After compression, the peak current is approximately 16~kA, the slice energy spread is 10~MeV, and the FWHM duration is 1.87~fs. The corresponding horizontal and vertical slice emittances at peak current are 0.36 mm·mrad and 0.30 mm·mrad, respectively. Due to the influence of the CSR, the compressed longitudinal phase spaces show distortions and increased energy spread in the high-current regions. Therefore, the choice of \( R_{56} \) values reflects a trade-off between achieving strong compression and mitigating excessive energy spread. The accelerator and undulator parameters corresponding to the head and tail current spike cases are summarized in Table~\ref{tab:SHINE}.

\begin{table}[h]
    \centering
    \caption{The AttoSHINE parameters for head-current spike case and tail-current spike case.}
    \label{tab:SHINE}
    \begin{tabular}{lccr}
        \hline\hline
        \textbf{Parameter} & \textbf{head} & \textbf{tail} & \textbf{Unit} \\
        \hline
        Bunch charge & 100 & 100 &  pC \\ 
        L1 rf phase & $-4.6$ & $-12.1$ & $\text{deg}$ \\   
        LH rf phase & $-142.2$ & $-152.0$ & $\text{deg}$ \\
        BC1 R56 & $-58.9$ & $-58.9$ & mm \\
        BC1 energy & $290$ & $280$ & MeV \\
        L2 rf phase & $-33.4$& $-28.9$ & $\text{deg}$ \\
        BC2 R56 & $-35.4$ & $-35.4$ & mm \\
        BC2 energy & $2.1$ & $2.2$ & GeV \\
        L3/L4 rf phase & $0$ & $0$ & $\text{deg}$ \\
        L4 energy & $8.7$ & $8.8$ & GeV \\        
        LTU1 R56 & $0.38$ & $0.61$ & mm \\        
        LTU1 horizontal dispersion & 0 & 0 & mm \\
        LTU1 vertical dispersion & 0 & 0 & mm \\
        FEL1 undulator period & $26$ & $26$ & mm \\
        FEL1 undulator segment length & $4$ & $4$ & m \\
        FEL1 photon energy & $6$ & $6$ & keV \\
        \hline\hline
    \end{tabular}
\end{table}

To evaluate the robustness of the attosecond generation under realistic operational conditions, we performed a jitter analysis reflecting expected CW operation at SHINE. At SHINE, the rms phase jitter is anticipated to be \(0.01^\circ\) for L1 and L2, and \(0.02^\circ\) for LH. The amplitude jitter is maintained below 0.01\% across all linac sections. 100 start-to-end simulations were conducted for each case, incorporating the specified jitter levels. The results, summarized in figure~\ref{fig:rms}, demonstrate that the self-chirping scheme remains effective under jitter conditions. Higher peak currents correlate with shorter FWHM durations of the current spike, albeit with a trade-off in the form of increased slice energy spread. Furthermore, the tail-current spike case shows a higher sensitivity to rf jitter than the head-current spike case.

\section{FEL simulation results}

\begin{figure}
\centering
\includegraphics[width=0.95\linewidth]{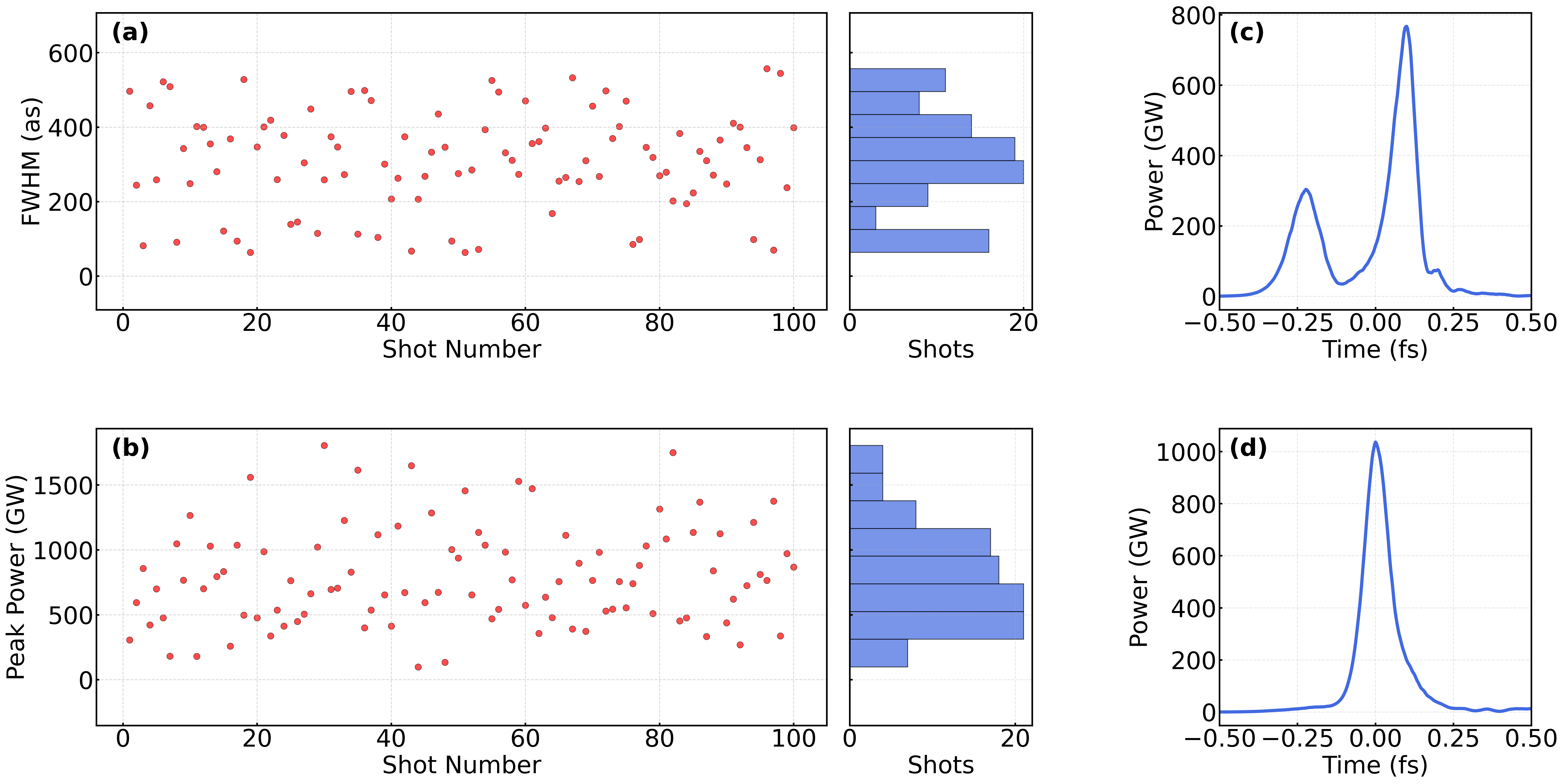}
\caption{FEL simulation results for the head-current-spike case at 6 keV including rf jitter effects. 
(a) FWHM pulse duration and (b) peak power for 100 different electron beams as a function of the beam index. 
(c) and (d) show temporal profiles of two representative single shots at the end of the 14th undulator. 
Panel (c) corresponds to a pulse with an envelope duration of 343~as and a peak power of 768~GW, while panel (d) shows a pulse with an envelope duration of 94~as and a peak power of 1037~GW. }
\label{fig:head_power}
\end{figure}

\begin{figure}
\centering
\includegraphics[width=0.95\linewidth]{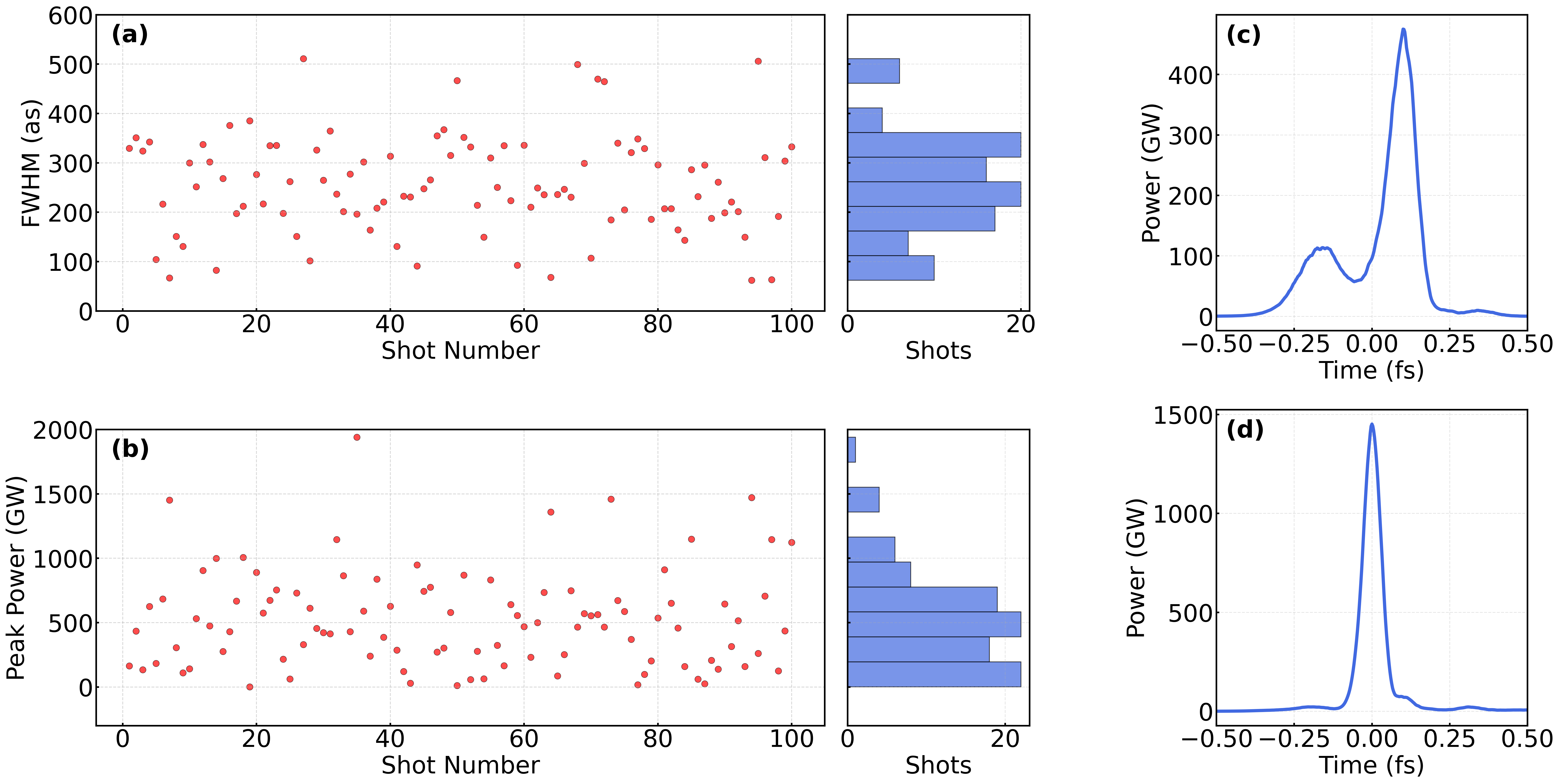}
\caption{ FEL simulation results for the tail-current-spike case at 6~keV including rf jitter effects. (a) FWHM pulse duration and (b) peak power for 100 different electron beams as a function of beam index. (c) and (d) show temporal profiles of two representative single shots at the end of the 14th undulator. (c) corresponds to a pulse with an envelope duration of 302~as and a peak power of 475~GW, while panel (d) shows a pulse with an envelope duration of 67~as and a peak power of 1451~GW.}
\label{fig:tail_power} 
\end{figure}

To evaluate the performance of the achievable attosecond XFEL pulses under realistic accelerator conditions, we performed FEL simulations using GENESIS 1.3 (version 4) \cite{genesis}. These simulations incorporate 100 statistically varied electron beam distributions that account for expected rf phase and amplitude jitter in the linac sections shown in figure~\ref{fig:rms}. The LSC effect of the electron beam within the undulator is included in the simulations. The results are shown in Figure~\ref{fig:head_power} and Figure~\ref{fig:tail_power}. These simulations are based on the SHINE FEL1 parameters \cite{cwshine2024}, employing an undulator with a period of 26 mm and a length of 4 m to generate XFEL pulses with a photon energy of 6 keV. For both cases, a slight reverse taper was incorporated into the undulator K values to match the chirp of the electron beam. Since the radiation generated outside the main current spike is negligible compared to that from the high-current region, only the spike region is shown in the figures. 

For the head-current spike case, with a taper of $\Delta K/K = 0.015\%$, the average FWHM pulse duration at the end of the fourteenth undulator is $308~\mathrm{as} \pm 133~\mathrm{as}$, and the peak power is $791~\mathrm{GW} \pm 381~\mathrm{GW}$, demonstrating that the target attosecond performance is well maintained despite the linac jitter. For the tail-current spike case, with a taper of $\Delta K/K = 0.024\%$, the average FWHM pulse duration is $255~\mathrm{as} \pm 100~\mathrm{as}$, and the peak power is $522~\mathrm{GW} \pm 376~\mathrm{GW}$. Here, the reported values are the mean over 100 samples. The quoted uncertainties correspond to one standard deviation across samples. Compared to head-current spike cases, tail-current spike cases are more sensitive to jitter and exhibit larger variations in beam parameters. Among the various cases considered under jitter conditions, some exhibit extreme bunch compression during transport through LTU1, degrading both the slice emittance and the slice energy spread. Consequently, the FEL gain is reduced in these cases, and the peak pulse power drops to a low level (a few gigawatts). All FWHM durations reported in this study refer to the envelope pulse duration, obtained via a global Gaussian fit to the radiation power profile. The head-current spike case produces, on average, both longer pulses and higher peak power, which can be attributed to its broader current profile and lower slice energy spread. Figure \ref{fig:head_power}c and \ref{fig:head_power}d present two typical pulses from the head-current spike case with a pulse duration of 343 as and 94 as, respectively. The left side of the figures corresponds to the head of the electron bunch. In both head-current and tail-current spike cases, the majority of pulses remain well below 600 as in duration. Furthermore, a fraction of the simulated shots exhibit peak powers exceeding 1~TW, accompanied by pulse durations below 100 as, as shown in figure \ref{fig:head_power}d and figure \ref{fig:tail_power}d. 

\begin{figure}
\centering
\includegraphics[width=0.8\linewidth]{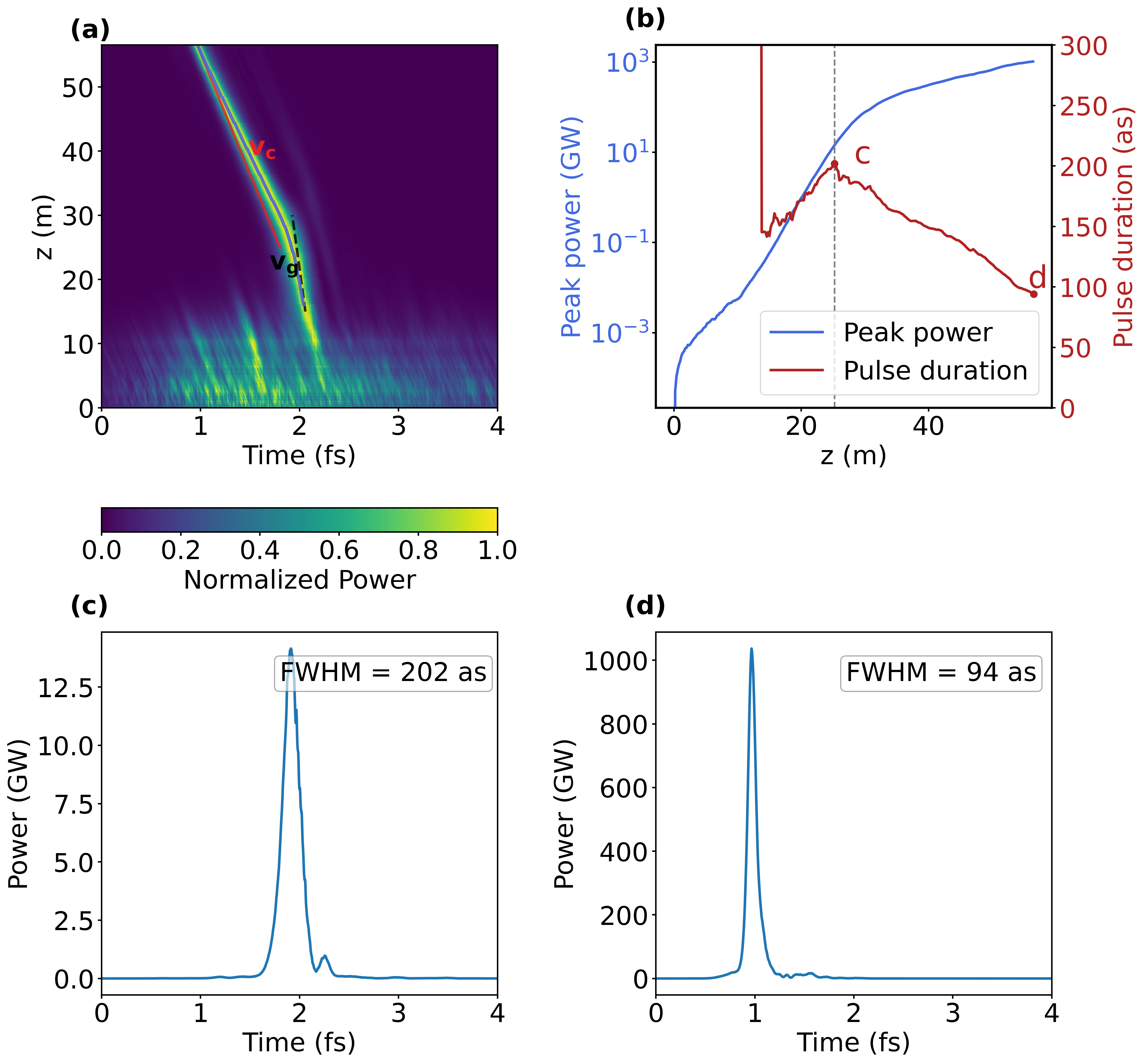}
\caption{Postsaturation behavior in the evolution of a terawatt attosecond pulse at 6 keV (selected shot shown in figure \ref{fig:head_power}d).
(a) Radiation power evolution along the undulator showing the emergence of a single attosecond X-ray pulse (profiles normalized independently at each $z$). The black dashed line indicates the group velocity $v_g=c\!\left(1-\tfrac{2\lambda_1}{3\lambda_u}\right)$, the red dashed line marks the speed of light $c$, and the blue ridge traces the pulse-peak trajectory; the color bar shows normalized power.(b) Peak power (blue, left axis) and FWHM pulse duration (red, right axis) versus undulator length $z$. Point $c$ marks the onset of pulse shortening; point $d$ denotes the undulator exit.(c,d) Temporal power profiles at positions $c$ and $d$, respectively, showing shortening from 202~as to 94~as while the peak power increases to the terawatt level.}
\label{fig:postsaturation} 
\end{figure}

We emphasize that the postsaturation regime is crucial for achieving higher-power, shorter attosecond XFEL pulses, an aspect rarely addressed in previous attosecond XFEL studies. To illustrate this, we focus on the shot shown in Figure \ref{fig:head_power}d for further analysis. As shown in figure \ref{fig:postsaturation}b, the peak-power growth departs from the exponential regime near point $c$, after which the pulse duration shrinks. From figure \ref{fig:postsaturation}a, the pulse group velocity drifts from $v_g = c\!\left(1-\tfrac{2\lambda_1}{3\lambda_u}\right)$~\cite{Bonifacio1994} (black dotted line), where $\lambda_1$ is the resonant FEL radiation wavelength and $\lambda_u$ is the undulator period, toward the speed of light (red solid line). This behavior in the postsaturation regime is similar to the superradiant behavior described in~\cite{bonifacio1989superradiance,bonifacio1990superradiant,bonifacio1991superradiant,Bonifacio1994,yang2020postsaturation}. In this regime, the strong radiation field drives a rapid phase-space rotation of the electrons, enabling efficient energy transfer from the beam to the leading edge of the spike, while the trailing edge returns energy to the electrons~\cite{kim2017synchrotron}. As a result, the peak power continues to rise and the pulse duration contracts, as illustrated by the temporal power profiles at positions $c$ and $d$ in Fig.~\ref{fig:postsaturation}(c,d). Although such superradiant behavior is often discussed for multi-stage XFEL concepts~\cite{PhysRevLett.114.244801,soft5}, we stress that it also arises in single-stage SASE XFEL. Recognizing and leveraging this regime is therefore central to optimizing attosecond XFEL performance. These observations highlight a critical design trade-off for attosecond XFELs. The current-spike width requires optimization rather than simple minimization. When the current spike becomes excessively short, slippage causes the XFEL pulse to outrun the electrons before reaching saturation, thereby limiting the achievable peak power. A key advantage of the self-chirping scheme, enabled by employing a high-charge electron beam, is that it can maintain sufficient temporal width while simultaneously achieving the high peak current necessary for high-power lasing.

\begin{figure}
\centering
\includegraphics[width=1\linewidth]{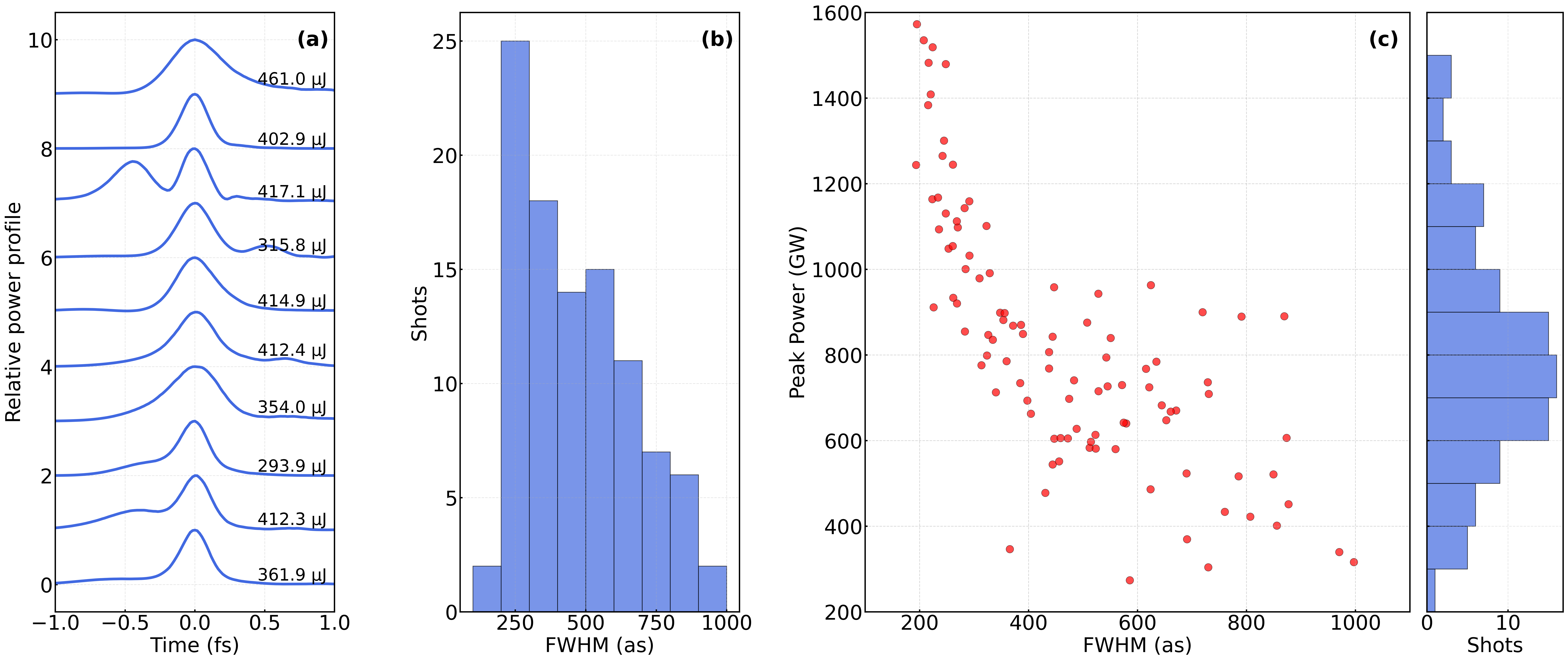}
\caption{ Soft X-ray FEL simulation results at 1~keV including rf jitter effects. 
(a) Temporal profiles of ten consecutive XFEL pulses. The corresponding pulse energy is indicated for each profile.
(b) Histogram of envelope FWHM duration obtained from 100 different electron beams.
(c) Scatter plot of peak power as a function of envelope FWHM duration.}
\label{fig:soft_head_jitter} 
\end{figure}

In the SHINE configuration, the LTU2 and LTU1 beamlines share similar lattice structures \cite{cwshine2024}, ensuring that the electron beam quality at the entrance of the FEL2 undulator is comparable to that at FEL1. This similarity permits the application of the self-chirping scheme at FEL2 as well, enabling the generation of high-power attosecond pulses in the soft X-ray regime. To demonstrate this capability, we retain the head-current spike beam parameters and replace the undulator with that of FEL2, which has a 55 mm period and a length of 4 m, to generate soft X-ray radiation at 1 keV. Figure \ref{fig:soft_head_jitter}a shows ten consecutive single-shot temporal profiles, while \ref{fig:soft_head_jitter}b and \ref{fig:soft_head_jitter}c present statistical results from 100 jittered electron beam shots. At the end of the eighth undulator segment, the average FWHM pulse duration is 469 as $\pm$ 202 as, and the average peak power is 824 GW $\pm$ 294 GW.

\section{Conclusion}

This work theoretically demonstrates that the self-chirping scheme enables generation of high-peak-power attosecond X-ray pulses at MHz repetition rates using a CW XFEL. We have systematically explained the universality of the self-chirping process by extending the self-chirping concept to the SHINE design framework and shown that it can deliver terawatt-class attosecond pulses in both the soft and hard X-ray regimes. By optimizing the quadrupole strengths in the switchyard section, we decoupled the \(R_{56}\) from the transverse dispersion for the bunch compression of the self-chirped beam. In contrast to the conventional scenario in which the current spike forms at the head of the bunch, we demonstrated that a tail current spike can be exploited to generate high power attosecond XFEL pulses. This is of significant value for various advanced XFEL operation modes, including multi-stage amplification \cite{wang2024millijoule,soft1}, attosecond XFEL carrying orbital angular momentum \cite{yan2023self,xu2024fel}, and enhanced self-seeding scheme \cite{hemsing2020enhanced,liu2025generation}. A systematic analysis was conducted on the amplitude and phase jitter expected from the linac sections under CW operation, and their impacts on the beam phase space were demonstrated. Simulation results show that at a photon energy of \(6\,\text{keV}\), it is possible to produce pulses with an average FWHM duration of approximately \(300\,\text{as}\) and average peak power of \(0.79\,\text{TW}\). When the photon energy is reduced to 1 keV, the scheme still produces attosecond pulses with average durations of around \(470\,\text{as}\) and average peak powers above \(0.8\,\text{TW}\). These results will provide guidance to conduct relevant experiments at SHINE in the future. 

Recent advances at SHINE, including dark-current suppression using an over-inserted plug in a normal-conducting VHF gun \cite{PhysRevAccelBeams.28.043401} and record high quality factors and accelerating gradients in 1.3 GHz superconducting cryomodules \cite{chen2025ultra}, suggest that a fully CW XFEL could ultimately deliver even brighter and shorter attosecond X-ray pulses than those predicted here for the current baseline configuration. Such a CW attosecond X-ray source would be transformative for ultrafast science, enabling MHz-rate pump–probe experiments \cite{chapman2025convergent, guo2024experimental} with unprecedented temporal resolution and statistical averaging. It would unlock real-time tracking of electron dynamics in correlated materials, quantum-coherent X-ray spectroscopy, and atomic-scale imaging of biological and chemical processes.

\section*{Acknowledgments}

The authors would like to acknowledge H. Y. and T. L. from SARI, G. G. from European XFEL, W. Q. from IHEP, M. G. and W. D. from DESY, I. I., T. H., and H. T. from SACLA for the fruitful discussions. 

\subsection*{Author Contributions} 
J. Y. and H. D. conceived the idea. B. Y. conducted the beam dynamics simulations. C. X. conducted the FEL simulations. All authors contributed equally to the writing of the manuscript. H. D. supervised this work. 

\subsection*{Funding}
This work was supported by the National Natural Science Foundation of China (12125508, 12541503, 12241501), and Shanghai Pilot Program for Basic Research – Chinese Academy of Sciences, Shanghai Branch (JCYJ-SHFY-2021-010). C.X. thanks the CAS-DAAD Joint Scholarship for its support. J. Y. and Y. C. acknowledge support from DESY (Hamburg, Germany), a member of the Helmholtz Association (HGF), and the European XFEL (Schenefeld, Germany).

\subsection*{Competing interests}
The authors declare that they have no competing interests.

\subsection*{Data Availability}
The data and simulation scripts that support the findings of this study can be obtained from the authors upon reasonable request.

\printbibliography

\end{document}